\documentclass[conference]{IEEEtran}

\IEEEoverridecommandlockouts
\usepackage{cite}
\usepackage{amsmath,amssymb,amsfonts}
\usepackage{algorithm}
\usepackage{algorithmic}
\usepackage{graphicx}
\usepackage{textcomp}
\usepackage{xcolor}
\usepackage{flushend}
\def\BibTeX{{\rm B\kern-.05em{\sc i\kern-.025em b}\kern-.08em
    T\kern-.1667em\lower.7ex\hbox{E}\kern-.125emX}}

\begin{document}

\title{Deep Reinforcement Learning-based Task Offloading in Satellite-Terrestrial Edge Computing Networks\\
}

\author{\IEEEauthorblockN{Dali Zhu\textsuperscript{1,2}, Haitao Liu\textsuperscript{1,2}, Ting Li\textsuperscript{1,2}, Jiyan Sun\textsuperscript{1}, Jie Liang\textsuperscript{1,2}, Hangsheng Zhang\textsuperscript{1,2}, Liru Geng\textsuperscript{1} and Yinlong Liu\textsuperscript{1,2,*}}
	\IEEEauthorblockA{\textsuperscript{1}\textit{Institute of Information Engineering, Chinese Academy of Sciences, Beijing, China} \\
		\textsuperscript{2}\textit{School of Cyber Security, University of Chinese Academy of Sciences, Beijing, China}\\
		\{zhudali, liuhaitao, liting0715, sunjiyan, liangjie, zhanghangsheng, gengliru,
		liuyinlong\}@iie.ac.cn}
	\textsuperscript{*}\textit{The corresponding author}\\
}

\maketitle

\begin{abstract}
In remote regions (e.g., mountain and desert), cellular networks are usually sparsely deployed or unavailable. With the appearance of new applications (e.g., industrial automation and environment monitoring) in remote regions, resource-constrained terminals become unable to meet the latency requirements. Meanwhile, offloading tasks to urban terrestrial cloud (TC) via satellite link will lead to high delay. To tackle above issues, Satellite Edge Computing architecture is proposed, i.e., users can offload computing tasks to visible satellites for executing. However, existing works are usually limited to offload tasks in pure satellite networks, and make offloading decisions based on the predefined models of users. Besides, the runtime consumption of existing algorithms is rather high.

In this paper, we study the task offloading problem in satellite-terrestrial edge computing networks, where tasks can be executed by satellite or urban TC. The proposed Deep Reinforcement learning-based Task Offloading (DRTO) algorithm can accelerate learning process by adjusting the number of candidate locations. In addition, offloading location and bandwidth allocation only depend on the current channel states. Simulation results show that DRTO achieves near-optimal offloading cost performance with much less runtime consumption, which is more suitable for satellite-terrestrial network with fast fading channel.
\end{abstract}


\section{Introduction}
With the emergence of 5G technology and the expansion of human activities, new applications such as industrial automation\cite{Luis_ICC_2020} and real-time environmental monitoring\cite{Liu_INFOCOM_2020}\cite{Wan_INFOCOM_2020} appear in remote regions. However, due to expensive construction and maintenance costs, cellular base stations are usually sparsely deployed or unavailable in remote regions\cite{Jia_WCNC_2020}. When resource-constrained terminals cannot meet the latency requirements of these new applications, computing tasks are offloaded to urban terrestrial cloud (TC)\cite{Guan_MobiCom_2020} for executing via satellites\cite{Lv_DataOffloading_2016} \cite{Zhang_EnergyEfficient_DataOffloading_2019}. However, the long propagation distance between remote terminals and urban TC will lead to high latency, which cannot meet the requirements of some delay-sensitive applications. Thanks to the emergence of low-earth-orbit (LEO) satellites, the propagation delay is significantly reduced. Furthermore, researchers proposed satellite edge computing (SatEC) architecture\cite{Xie_Network_2020}\cite{Yan_SatEC_2019}\cite{Wang_SatEC_IoT_2019} by referring to mobile edge computing (MEC)\cite{MEC_survey}. Remote terminals can directly offload computing tasks to nearby visible satellites for executing, which further reduces the offloading delay.

\begin{figure}[htbp]
	\centerline{\includegraphics[width = 0.45\textwidth]{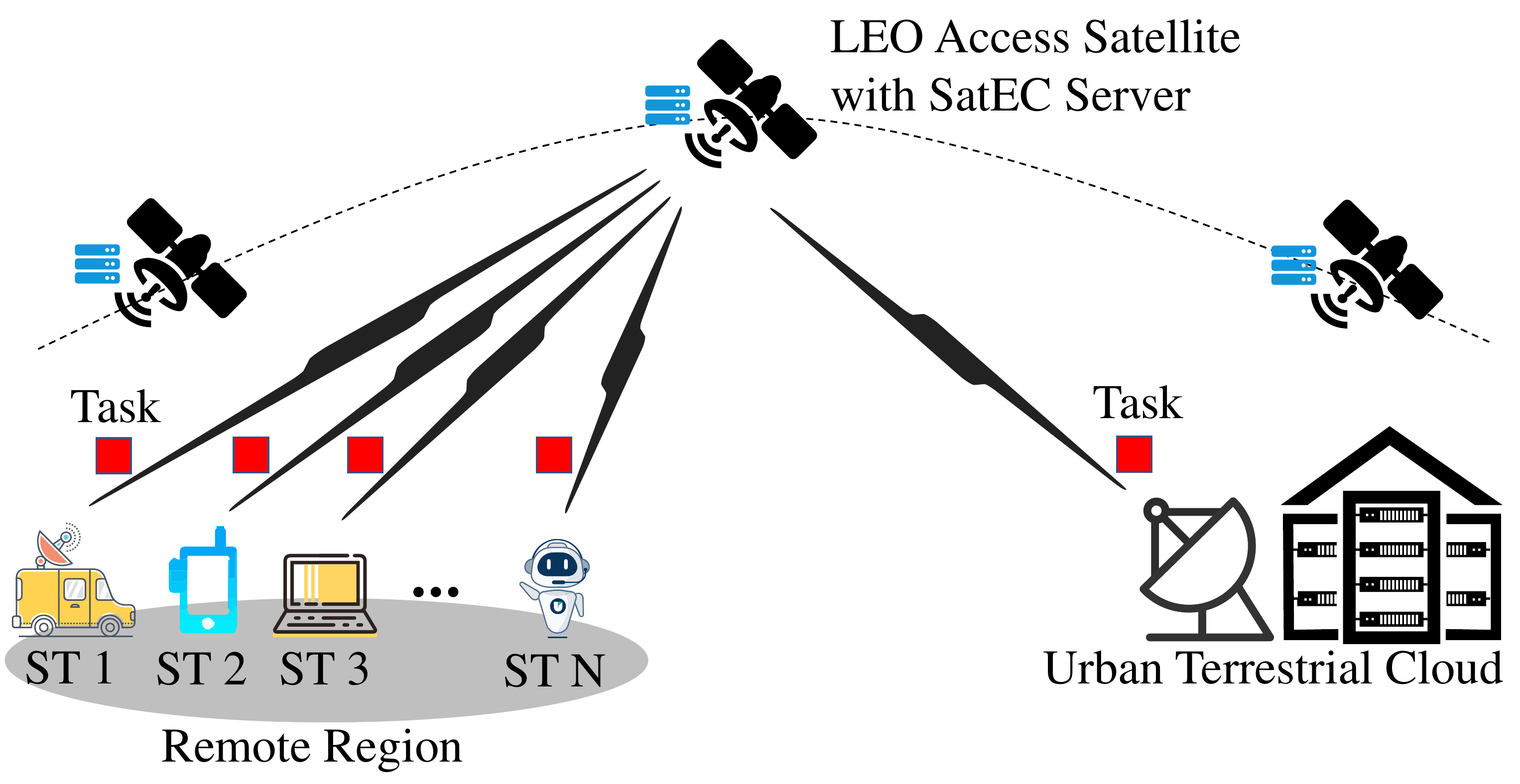}}
	\caption{Task Offloading in Satellite-Terrestrial Edge Computing Networks}
	\label{system_model}
\end{figure}

Recently, there are several efforts focusing on task offloading in SatEC networks. Zhang et al. \cite{Zhang_SAIC_2020} proposed a satellite-aerial integrated computing architecture, where ground/aerial users offload tasks to high-altitude platforms or LEO satellites. Considering the intermittent communication caused by satellite orbiting, Wang et al. \cite{Wang_Game_2020} proposed a IoT-to-Satellite offloading method based on game theory. However, they do not consider the cooperation between SatEC server and urban terrestrial data centers. Actually, due to the limited computing capacity and energy reservation of satellite, when a large number of tasks are simultaneously offloaded, SatEC servers need to cooperate with urban TC to provide satisfying computing service. As shown in Fig. \ref{system_model}, in a satellite-terrestrial integrated network, the LEO access satellite can choose to locally execute the offloaded tasks, or transparently forward them to its connected urban TC. 

Furthermore, although some existing works focus on offloading in satellite-terrestrial integrated network, they require some predefined models. For examples, the flight trajectories of aerial users are required in \cite{Zhang_SAIC_2020}, and the flight trajectory of unmanned aerial vehicle is required in \cite{Zhou_GLOBECOM_2019}, which are usually difficult to obtain in practice. Instead, we propose to make offloading decisions only based on current channel states, which is more convenient to obtain. In addition, to optimize the delay and energy consumption in SatEC network, researchers usually formulate the offloading decision and bandwidth allocation problem as a mixed-integer programming (MIP) problem\cite{Kim_INFOCOM_2020}\cite{Huang_ACCESS_2020}. The 3D hypergraph matching \cite{Zhang_SAIC_2020}, game-theoretic approach \cite{Wang_Game_2020} and a multiple-satellite offloading method \cite{Gao_GLOBECOM_2019} have been proposed to solve the hard MIP problem. However, both of them require considerable number of iterations to reach a satisfying optimum. Hence, they are not suitable for making real-time offloading decisions, especially under the fast fading channels\cite{Espinosa_GLOBECOM_2019} caused by high speed movement of LEO satellites\cite{Maattanen_GLOBECOM_2019}.

In this paper, we consider a satellite-terrestrial edge computing network and model the offloading cost as weighted sum of latency and energy consumption. To minimize the offloading cost, the offloading location decision and bandwidth allocation is formulated as a MIP problem. Then, we propose a low-complexity Deep Reinforcement learning-based Task Offloading (DRTO) algorithm to solve it. Specifically, the deep neural network (DNN)\cite{Jeong_SoCC_2018} only takes the current channel states as inputs, and outputs a relaxed offloading location, which is then quantized into a set of candidate binary offloading locations. Given a candidate location, a bandwidth allocation convex problem is solved by CVXPY \cite{cvxpy_home} tool. The main contributions of this paper are summarized as follows:

\begin{itemize}
	\item \textbf{Satellite-terrestrial cooperative offloading.} We consider a satellite-terrestrial cooperative edge computing architecture, where the tasks can be executed by either SatEC server or urban TC. The offloading location decision and bandwidth allocation is formulated as a MIP problem.
	\item \textbf{Model-free learning.} The proposed DRTO algorithm makes offloading decision only based on the current channel states. Meanwhile, DRTO can improve its offloading policy by learning from the real-time trend of channel states, which adapts to the high dynamics of satellite-terrestrial networks.
	\item \textbf{Low time complexity.} Compared with traditional optimization methods, DRTO completely removes the need of solving hard MIP problem. Furthermore, we dynamically adjust the size of action space to speed up the learning process. Simulation results show that the runtime consumption of DRTO is significantly decreased, while the offloading cost performance is not compromised.
\end{itemize}

The rest of this paper is organized as follows: We describe the system model and formulates the offloading cost minimization problem in Section \ref{system_model_and_problem_formulation}. The details of DRTO algorithm is introduced in Section \ref{DRTO}. In Section \ref{simulation_results}, simulation results are presented. Finally, the paper is concluded in Section \ref{conclusion}.

\section{System Model and Problem Formulation}

\label{system_model_and_problem_formulation}

As shown in Fig. \ref{system_model}, LEO satellites fly above the surface of earth at high speed, and connect the remote STs to ground station. TC is directly connected to ground station via an optical fiber and its transmission delay can be ignored. We assume that the access satellite is always available, and consider $N$ STs denoted by $\mathcal{N}=\{1,2,...,N\}$ and a TC within the coverage of the same access satellite. For simplicity, we denote the wireless signal traveling from ST to its access satellite as the 1st-hop, and the 2nd-hop from access satellite to TC. We assume the access satellite can measure channel states before deciding the offloading locations and allocating the bandwidth. The notations used throughout the paper are list in Table \ref{tab_notations}.

\begin{table}[htbp]
	\caption{Notations Used in this Paper}
	\begin{center}
		\begin{tabular}{|c|c|}
			\hline
			\textbf{Notation} & \textbf{Description}\\ 
			\hline
			$x_n$ & Offloading location of $n$-th ST\\
			\hline
			$\alpha_nB$ & Bandwidth allocated for $n$-th ST\\
			\hline
			$\alpha_{N + n}B$ & Bandwidth allocated for forwarding the task of $n$-th ST\\
			\hline
			$B$ & Total bandwidth of access satellite\\
			\hline
			$p_n$ & Transmission power of $n$-th ST\\
			\hline
			$p_{SAT}$ & Transmission power of access satellite\\
			\hline
			$h_n$ & Channel gain between $n$-th ST and its access satellite\\
			\hline
			$h_{TC}$ & Channel gain between access satellite and TC\\
			\hline
			$N_0$ & Noise power at the receiver\\
			\hline
			$L$ & Size of task\\
			\hline
			$k$ & Computational intensity\\
			\hline
			$f_1$ & CPU frequency of SatEC server\\
			\hline
			$f_0$ & CPU frequency of TC\\
			\hline
			$p_c$ & Computing Power Consumption of SatEC server\\
			\hline
			$\lambda$ & Latency-Energy Weight Parameter\\
			\hline
		\end{tabular}
		\label{tab_notations}
	\end{center}
\end{table}

\subsection{Offloading Location}

For the task offloaded by $n$-th ST, its access satellite can choose to locally process or transparently forward to its connected TC. We denote the offloading location of $n$-th ST as $x_n$, where $x_n=1$ and $x_n=0$ respectively denotes SatEC server and TC.

\subsection{Offloading Cost}

The quality of service (QoS) mainly depends on user-perceived latency and energy consumption. Moreover, considering the precious energy reservation of satellites, we also include the energy consumption of satellites into cost. The detailed definitions of offloading cost for different locations are given as follows:

\subsubsection{Offloaded to SatEC server}

When tasks are offloaded to SatEC server, the cost mainly consists of STs' transmission cost and SatEC server's computing cost. We denote $\alpha_n$ as the proportion of bandwidth allocated for $n$-th ST, then the $n$-th ST's 1-st hop transmission rate is given by $C_{1, n} = \alpha_nB\log_2\left(1 + p_nh_n/N_0\right)$, where $B$ denotes the total bandwidth of access satellite, $p_n$ denotes the transmission power of $n$-th ST, $h_n$ denotes the channel gain between $n$-th ST and its access satellite, and $N_0$ denotes the noise power at the receiver.

Based on the 1st-hop transmission rate $C_{1, n}$, the transmission latency is given by $T_{1, n} = L/C_{1, n}$, where $L$ denotes the task size (in bits). Then, the energy consumed by $n$-th ST for transmission is given by $E_{1, n} = p_nT_{1, n}$.

We simply ignore the queuing delay. The computing latency at SatEC server is given by $T_{1, n}^c = kL/f_1$, where $k$ denotes the computational intensity (in cycles/bit) of task, and $f_1$ denotes the CPU frequency (in cycles/s) of SatEC server. The energy consumed by SatEC server for computing is given by $E_{1, n}^c = p_cT_{1, n}^c$, where $p_c$ denotes the computing power consumption (in Watt) of SatEC server. 

Therefore, the total latency that $n$-th ST perceived and energy consumed for $n$-th ST are respectively given by $T_n^{SAT}=T_{1,n}+T_{1,n}^c$ and $E_n^{SAT}=E_{1,n}+E_{1,n}^c$.

\subsubsection{Offloaded to TC}

When tasks are offloaded to TC, apart from the transmission cost of STs, the forwarding cost of access satellite and computing cost of TC should be included. We denote $\alpha_{N + n}$ as the proportion of bandwidth allocated for forwarding the task of $n$-th ST, then the 2nd-hop transmission rate for $n$-th ST is given by $C_{2, n} = \alpha_{N + n}B\log_2\left(1 + p_{SAT}h_{TC}/N_0\right)$, where $p_{SAT}$ denotes the transmission power of access satellite, $h_{TC}$ denotes the channel gain between access satellite and TC. Therefore, the forwarding latency and energy consumption for $n$-th ST are respectively given by $T_{2, n} = L/C_{2, n}$ and $E_{2, n} = p_{SAT}T_{2, n}$.

The computing latency at TC is given by $T_{2, n}^c = kL/f_0$, where $f_0$ denotes the CPU frequency (in cycles/s) of TC. Thanks to the continuous electrical power supply for TC, we simply ignore the computing energy consumption of TC. Therefore, the total latency that $n$-th ST perceived and energy consumed for $n$-th ST are respectively given by $T_n^{TC}=T_{1,n}+T_{2,n}+T_{2,n}^c$ and $E_n^{TC}=E_{1,n}+E_{2,n}$.

\subsection{Problem Formulation}

As mentioned above, the offloading cost is mainly composed of latency and energy consumption, which depends on offloading locations, current channel states and bandwidth allocation. Therefore, the offloading cost minimization problem $\mathcal{P}$ is formulated as following:
\begin{subequations}
	\begin{equation}
	\begin{split}
	\mathcal{P}:\min_{\boldsymbol{x}, \boldsymbol{\alpha}} F\left(\boldsymbol{x}, \boldsymbol{\alpha}\right)=\sum_{n = 1}^Nx_n\left[\lambda T_n^{SAT} + \left(1 - \lambda\right)E_n^{SAT}\right] \\+\left(1 - x_n\right)\left[\lambda T_n^{TC} + \left(1 - \lambda\right)E_n^{TC}\right]
	\end{split}
	\end{equation}
	\begin{equation}
	\begin{aligned}
	&s.t.&x_n \in \{0, 1\}, \forall{n}\in\mathcal{N}
	\end{aligned}
	\end{equation}
	\begin{equation}
	\begin{aligned}
	& &0\leq\sum_{n = 1}^{2N}\alpha_n\leq 1
	\end{aligned}
	\end{equation}
	\begin{equation}
	\label{bw_pos_cons}
	\begin{aligned}
	& &\alpha_n\geq 0,\forall{n}\in\{1,...,2N\}
	\end{aligned}
	\end{equation}
\end{subequations}
where $\lambda$ denotes the weight parameter for balancing the latency and energy consumption.

It can be seen that problem $\mathcal{P}$ is a mixed-integer programming problem, in which the $0-1$ integer variable $\boldsymbol{x}$ and the continuous variable $\boldsymbol{\alpha}$ are mutually coupled. This problem is commonly reformulated by specific relaxation approach and then solved by powerful convex optimization techniques. However, these methods perform considerable iterations, and the original problem cannot be solved within channel coherent time, especially when many STs simultaneously offload tasks. To tackle this dilemma, we are motivated to propose a effective low-complexity Deep Reinforcement learning-based Task Offloading algorithm to obtain the near-optimal solution. Specifically, we adopt a DNN to map the current channel states to offloading locations, and improve the DNN via reinforcement learning.

\section{DRTO: Deep Reinforcement Learning for Task Offloading}
\label{DRTO}

To minimize the offloading cost, we design an offloading algorithm $\pi:\boldsymbol{h}\xrightarrow{}\boldsymbol{x}^*$ that quickly selects the optimal offloading location $\boldsymbol{x}^*=[x_1^*,x_2^*,...,x_N^*]$ only based on the current channel state $\boldsymbol{h}=[h_1,h_2,...,h_N,h_{TC}]$.

\begin{figure}[htbp]
	\centerline{\includegraphics[width =0.45\textwidth]{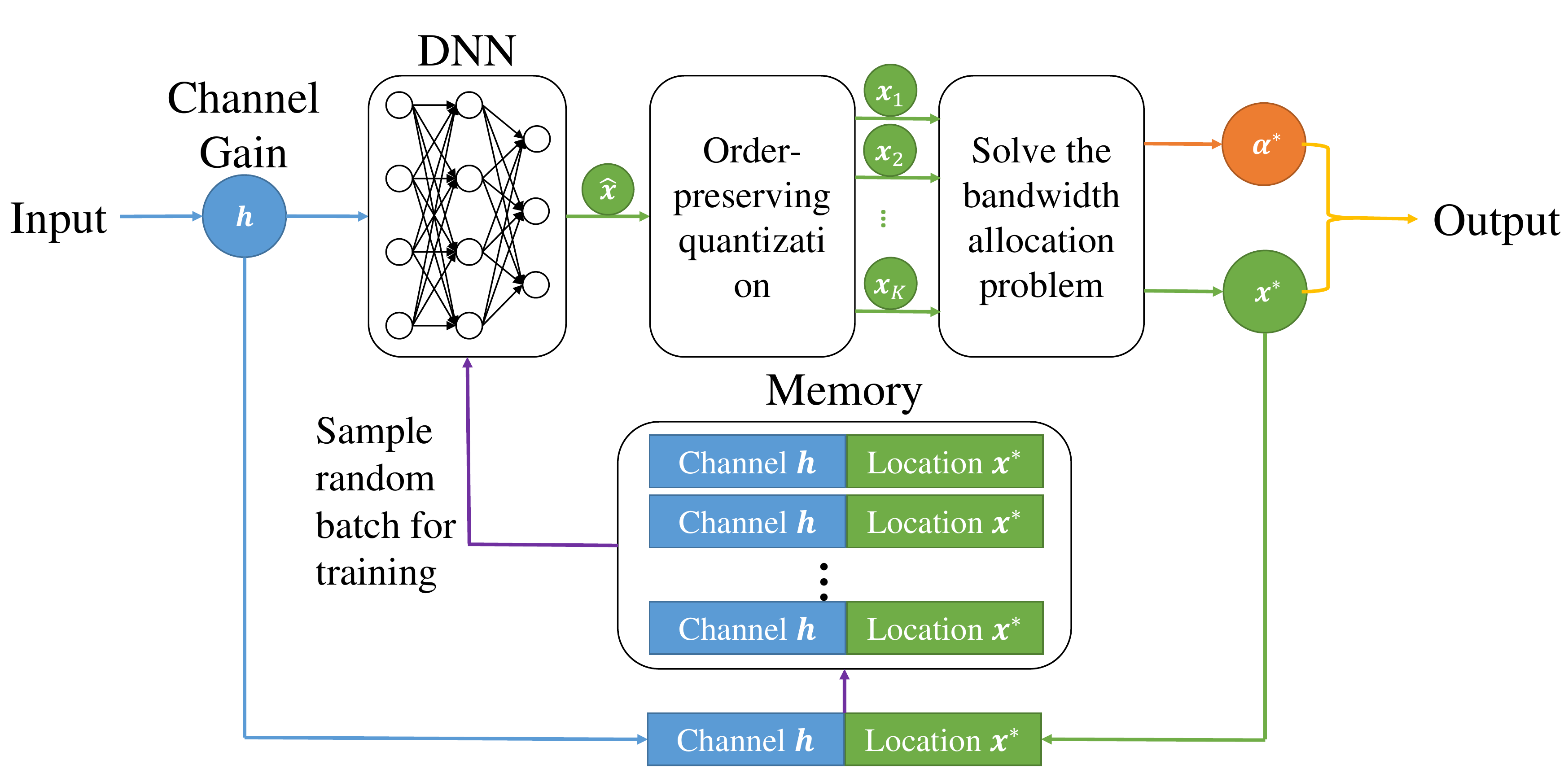}}
	\caption{The diagram of DRTO}
	\label{DRTO_structure}
\end{figure}

The diagram of DRTO is shown in Fig. \ref{DRTO_structure}. First, the DNN takes the current channel gain $\boldsymbol{h}$ as inputs, and generates a relaxed offloading location $\boldsymbol{\hat{x}}$. Then, we quantize the relaxed location $\boldsymbol{\hat{x}}$ into $K$ candidate binary offloading locations, namely $\boldsymbol{x}_1, \boldsymbol{x}_2,...,\boldsymbol{x}_K$. The optimal location $\boldsymbol{x}^*$ is obtained by solving a series of bandwidth allocation convex problems. Subsequently, the newly obtained channel state-offloading location pair $(\boldsymbol{h}, \boldsymbol{x}^*)$ is added into replay memory. A random batch will be sampled from memory to improve the DNN every $\delta$ time frames. To further reduce the runtime consumption, we dynamically adjust $K$ to speed up the learning process. In the following subsections, the details of above stages are described. The pseudocode of DRTO algorithm is summarized in Algorithm 1. 

\begin{algorithm}[h]
	\caption{The DRTO Algorithm}
	\begin{algorithmic}[1]
		\STATE \textbf{Input:} Current channel gain $\boldsymbol{h}$.
		\STATE \textbf{Output:} Optimal Offloading location $\boldsymbol{x}^*$ and corresponding bandwidth allocation $\boldsymbol{\alpha}^*$.
		\FOR{$t=1,2,...,T$}
		\STATE DNN generates a relaxed offloading location $\hat{\boldsymbol{x}}$.
		\STATE Quantize $\hat{\boldsymbol{x}}$ into $K_t$ candidate binary offloading locations $\boldsymbol{x}_k, k\in{1,2,...,K_t}$.
		\FOR{$k=1,2,...,K_t$}
		\STATE Given binary offloading location $\boldsymbol{x}_k$, the bandwidth allocation $\boldsymbol{\alpha}_{\boldsymbol{x}_k}$ and offloading cost $F\left(\boldsymbol{x}_k,\boldsymbol{\alpha}_{\boldsymbol{x}_k}\right)$ are obtained by solving $\mathcal{P}'$.
		\ENDFOR
		\STATE Obtain the optimal offloading location with respect to $\boldsymbol{x}^*=\arg\min_{\boldsymbol{x}_k,k\in{1,2,...,K}}F\left(\boldsymbol{x}_k,\boldsymbol{\alpha_{\boldsymbol{x}_k}}\right)$.
		\STATE Add newly obtained channel state-offloading location pair $\left(\boldsymbol{h}_t,\boldsymbol{x}^*\right)$ into replay memory.
		\IF{$t\bmod\delta==0$}
		\STATE Sample a random batch from memory for training DNN.
		\ENDIF
		\IF{$t\bmod\Delta==0$}
		\STATE Adjust $K_t$ using $(\ref{adjust_K})$.
		\ENDIF
		\ENDFOR
	\end{algorithmic}
\end{algorithm}
\subsection{Generate the Offloading Location}
As shown in the upper part of Fig. \ref{DRTO_structure}, in each time frame, the fully connected DNN takes the current channel gain $\boldsymbol{h}$ as inputs, and generates a relaxed offloading location $\hat{\boldsymbol{x}}=[\hat{x}_1,\hat{x}_2,...,\hat{x}_N]$ (each entry is relaxed into $[0,1]$ interval). Then, the relaxed location $\hat{\boldsymbol{x}}$ is quantized into $K$ binary locations. Given a candidate location $\boldsymbol{x}_k$, DRTO solves a bandwidth allocation convex problem, and obtains the offloading cost. Subsequently, the optimal offloading location $\boldsymbol{x}^*$ is selected according to the minimal offloading cost.

Although the mapping from channel state to offloading location is unknown and complex, thanks to the universal approximation theorem \cite{marsland2015machine}, we adopt a fully connected DNN to approximate this mapping. The DNN is characterized by the weights that connect the hidden neurons, and composed of four layers, namely input layer, two hidden layers and ouput layer. Here, we respectively use ReLU and sigmoid activation function in the hidden layers and output layer, thus each entry of the output relaxed offloading location satisfies $\hat{x}_n\in(0,1)$.

Then, the $\hat{\boldsymbol{x}}$ is quantized into $K$ candidate binary offloading locations, where $K\in[1,2^N]$. Intuitively, a larger $K$ creates higher diversity in the candidate offloading location set, thus increasing the chance of finding the global optimal offloading location, but resulting in higher computational complexity. We adopt an order-preserving quantization method proposed in \cite{Huang_DROO_2020} for the trade-off of performance and complexity. In order-preserving quantization, the $K$ is relatively small, but the diversity of candidate offloading locations is guaranteed. Its main idea is preserving the order when quantization, i.e., for each quantized location $\boldsymbol{x}_k=[x_{k,1},x_{k,2},...,x_{k,N}]$, $x_{k,n}\leq x_{k,m}$ should be held if $\hat{x}_{n}\leq \hat{x}_{m}$ for all $n,m\in\{1,2,...,N\}$. Specifically, a series of $K$ quantized locations $\{\boldsymbol{x}_k\}$ are generated as following:

1) Each entry of the 1st binary offloading location $\boldsymbol{x}_1$ is given by
\begin{equation}
x_{1,n}=
\begin{cases}
1&\hat{x}_n>0.5,\\
0&\hat{x}_n\leq0.5.
\end{cases}
n=1,2,...,N
\end{equation}

2) As for the remaining $K-1$ offloading locations, we first sort each entry of $\hat{\boldsymbol{x}}$ according to their distance to 0.5, i.e., $|\hat{x}_{(1)}-0.5|\leq|\hat{x}_{(2)}-0.5|\leq...\leq|\hat{x}_{(N)}-0.5|$, where $\hat{x}_{(n)}$ denotes the sorted $n$-th entry. Hence, each entry of the $k$-th offloading location $\boldsymbol{x}_k,k=2,3,...,K$ is given by
\begin{equation}
x_{k,n}=
\begin{cases}
1&\hat{x}_n>\hat{x}_{(k-1)},\\
1&\hat{x}_n=\hat{x}_{(k-1)}\text{ and }\hat{x}_{(k-1)}\leq 0.5,\\
0&\hat{x}_n=\hat{x}_{(k-1)}\text{ and }\hat{x}_{(k-1)}>0.5,\\
0&\hat{x}_n<\hat{x}_{(k-1)}.\\
\end{cases}
n=1,2,...,N
\end{equation}

Here we obtain $K$ candidate offloading locations, given a candidate offloading location $\boldsymbol{x}_k$, the original offloading cost minimization problem $\mathcal{P}$ is transformed into a convex problem on $\boldsymbol{\alpha}$
\begin{subequations}
	\begin{equation}
	\mathcal{P'}:\min_{\boldsymbol{\alpha}} F\left(\boldsymbol{x}_k,\boldsymbol{\alpha}\right)
	\end{equation}
	\begin{equation}
	\begin{aligned}
	&s.t.&0\leq\sum_{n = 1}^{2N}\alpha_n\leq 1
	\end{aligned}
	\end{equation}
\end{subequations}
which can be solved by convex optimization tool like CVXPY \cite{cvxpy_home}. Then we obtain the optimal bandwidth allocation $\boldsymbol{\alpha}^*_{\boldsymbol{x}_k}$ and minimum offloading cost $F\left(\boldsymbol{x}_k,\boldsymbol{\alpha}^*_{\boldsymbol{x}_k}\right)$ with the given candidate offloading location $\boldsymbol{x}_k$. By repeatedly solving the problem $\mathcal{P'}$ for each candidate offloading location, the best offloading location is selected by
\begin{equation}
\boldsymbol{x}^*=\arg\min_{\{\boldsymbol{x_k}\},k=1,2,...K}F\left(\boldsymbol{x}_k,\boldsymbol{\alpha}^*_{\boldsymbol{x}_k}\right)
\end{equation}
along with its corresponding optimal bandwidth allocation $\boldsymbol{\alpha}^*$.
\subsection{Update the Offloading Policy}
Due to the rapid changes of satellite-terrestrial channel states, in order to reduce the offloading cost, the offloading policy should be updated in time. Different from traditional deep learning, the training samples of DRTO are composed of the latest channel state $\boldsymbol{h}$ and offloading location $\boldsymbol{x}^*$. Since the current offloading location is generated according to the policy in the last time frame, the training samples in adjacent time frames are strongly correlated. If the latest samples are used to train the DNN immediately, the network will be updated in an inefficient way, and the offloading policy even may not converge. Thanks to the experience replay mechanism \cite{Atari} proposed by Google DeepMind, the newly obtained state-location pair $\left(\boldsymbol{h},\boldsymbol{x}^*\right)$ is added to the replay memory, and replaces the oldest one if the memory is full. Subsequently, a random batch are sampled from the memory to improve the DNN. The cross-entropy loss is reduced by utilizing the Adam optimizer \cite{Kingma_2014}. Such iterations repeat and the policy of the DNN is gradually improved.

By utilizing the experience replay mechanism, we construct a dynamic training dataset for DNN. Thanks to the random sampling, the convergence is fastened because the correlation between training samples is reduced. Since the memory space is finite, the DNN is updated only according to the recent experience, and the offloading policy $\pi$ is always adapted to the recent channel changes.

\subsection{Dynamically Adjust $K$}
For each candidate offloading location, a bandwidth allocation convex problem is solved. Intuitively, a larger $K$ can lead to a better temporary offloading decision and a better long-term offloading policy. However, to select the optimal offloading location $\boldsymbol{x}^*$ in each time frame, repeatedly solving bandwidth allocation problem $(\mathcal{P'})$ $K$ times leads to high computational complexity. Therefore, there exists a trade-off between performance and complexity according to the setting of $K$.
\begin{figure}[htbp]
	\centerline{\includegraphics[width =0.45\textwidth]{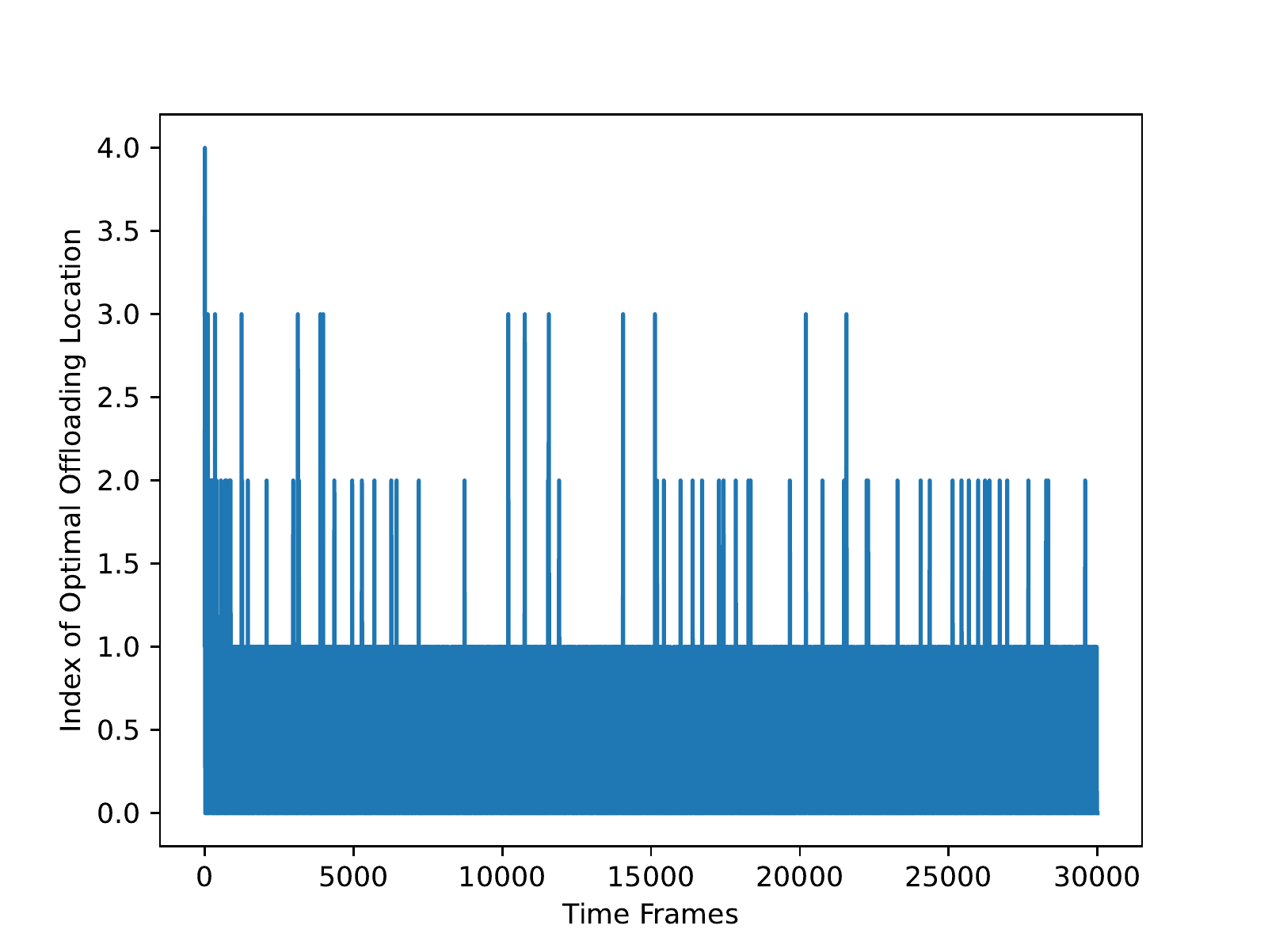}}
	\caption{The index of optimal offloading location with $K=N=5$}
	\label{idx_best}
\end{figure}

With a fixed $K=N$, we plot the index of optimal offloading location in each time frame. As shown in Fig. \ref{idx_best}, at the very beginning of the learning process, the index of the optimal offloading location is relatively large. As the offloading policy improves, we observe that most of the optimal offloading location are the first location generated by above order-preserving quantization method. This indicates that a large value of $K$ is computationally inefficient and unnecessary. In other words, most of the quantized offloading location in each time frame are redundant. Therefore, to speed up the algorithm, we can gradually adjust $K$, and the performance will not be compromised.

We denote $K_t$ as the number of quantized offloading locations at time frame $t$. Inspired by \cite{Huang_DROO_2020}, we initially set $K_1 = N$. For every $\Delta$ time frames, $K_t$ will be adjusted once. In an adjustment time frame, to increase the diversity of candidate offloading locations, $K_t$ is tuned to $\max\left(k_{t-1}^*,...,k_{t-\Delta}^*\right)+1$. Therefore, $K_t$ is given by
\begin{equation}
\label{adjust_K}
K_t=
\begin{cases}
N&t=1,\\
\min\left(\max\left(k_{t-1}^*,...,k_{t-\Delta}^*\right)+1,N\right)&t\text{ mod }\Delta=0,\\
K_{t-1}&\text{otherwise}.
\end{cases}
\end{equation}
\section{Simulation Results}
\label{simulation_results}

In this section, the performance of the proposed DRTO algorithm is evaluated via simulations. The average channel gain $h_n$ or $h_{TC}$ follows the free space path loss model
\begin{equation}
h=A_d\left(\frac{c}{4\pi f_cd}\right)^{d_e}.
\end{equation}
The first and second hidden layer of DNN have $120$ and $80$ hidden neurons, respectively. The initial parameters of the DNN follow a normal distribution with zero-mean. The DRTO algorithm is implemented in Python with TensorFlow 2.0. We respectively evaluate the performance of convergence, offloading cost and runtime. Other default parameters are listed in Tab. \ref{tab_simulation_parameters}.


\begin{table}[htbp]
	\caption{Simulation Parameters Setup}
	\begin{center}
		\begin{tabular}{|c|c|}
			\hline
			\textbf{Parameters} & \textbf{Value}\\ 
			\hline
			Transmission power of ST $p_n$ and satellite $p_{SAT}$ (W)& 1, 3\\
			\hline
			Antenna gain $A_d$ and path loss exponent $d_e$ & 4.11, 2.8\\
			\hline
			Carrier frequency $f_c$ (GHz)& 30\\
			\hline
			Total bandwidth $B$ (MHz)& 800\\
			\hline
			Receiver noise power $N_0$ (W)& $10^{-9}$\\
			\hline
			Task size $L$ (MB)& 100\\
			\hline
			Computational intensity $k$ (cycles/bit)& 10\\
			\hline
			Computing Power Consumption of SatEC server $p_c$ (W)& 0.5\\
			\hline
			CPU frequency of SatEC server $f_1$ and TC $f_0$ (GHz)& 0.4, 3\\
			\hline
			Latency-energy weight parameter $\lambda$ & 0.5\\
			\hline
			Training interval $\delta$ & 10\\
			\hline
			Random batch size & 128\\
			\hline
			Replay memory size & 1024\\
			\hline
			Learning rate & 0.01\\
			\hline
		\end{tabular}
		\label{tab_simulation_parameters}
	\end{center}
\end{table}

\subsection{The Performance of Convergence}
\begin{figure}[htbp]
	\centerline{\includegraphics[width =0.45\textwidth]{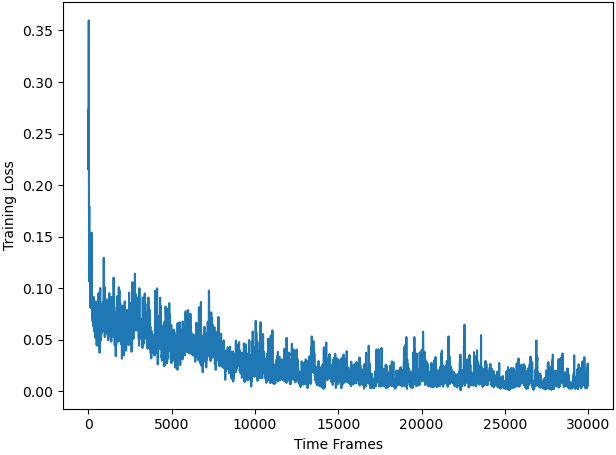}}
	\caption{The traning loss of DRTO}
	\label{DRTO_training_loss}
\end{figure}

The DRTO algorithm is evaluated over $30000$ time frames. In Fig. \ref{DRTO_training_loss}, we plot the training loss of the DNN, which gradually decreases and stabilizes at around $0.02$, whose fluctuation is mainly due to the random sampling of training data.

\begin{figure}[htbp]
	\centerline{\includegraphics[width =0.45\textwidth]{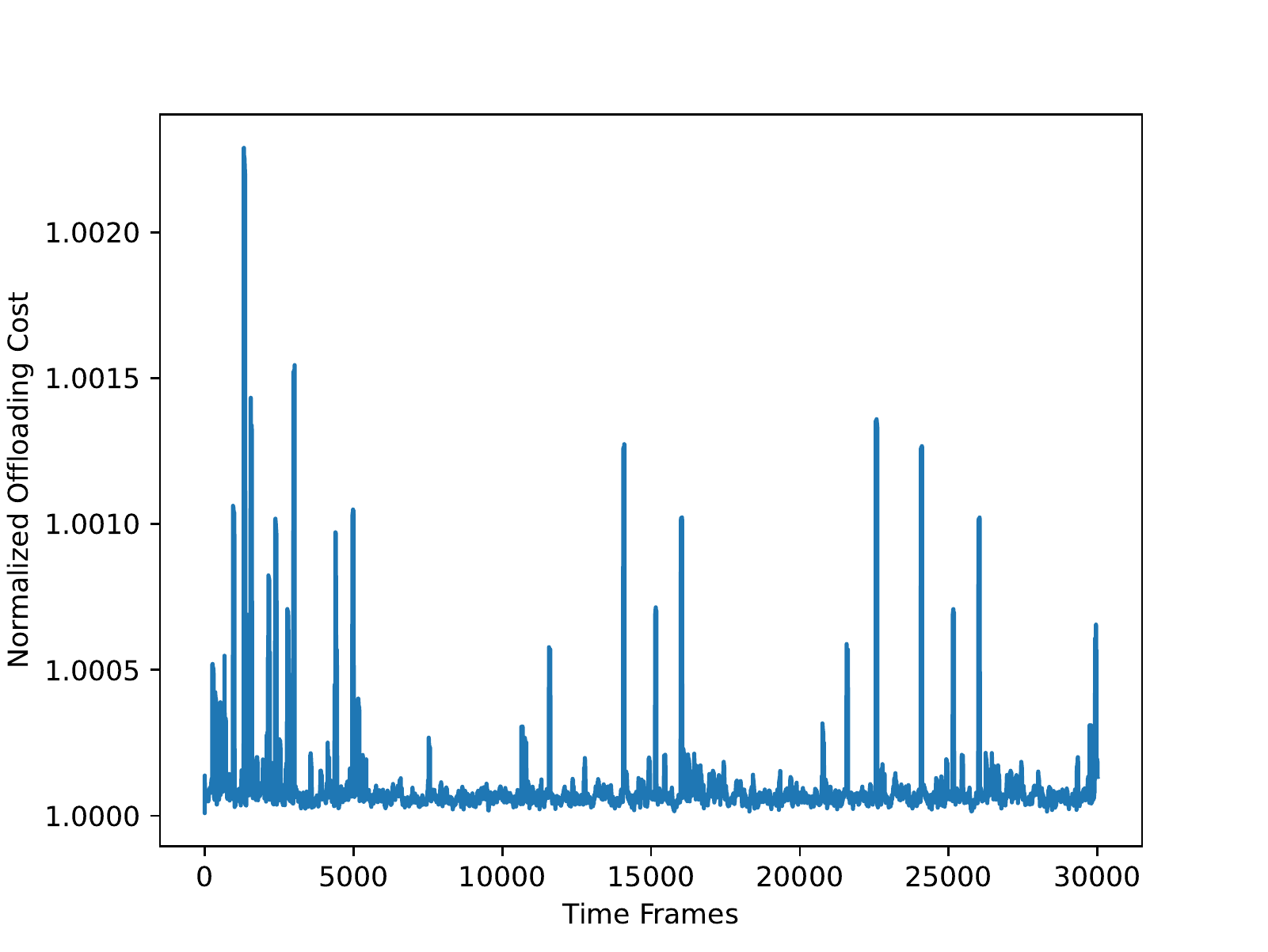}}
	\caption{Normalized offloading cost with $\Delta=64$}
	\label{normalized_offloading_cost}
\end{figure}

In Fig. \ref{normalized_offloading_cost}, we plot the normalized offloading cost, which is defined as
\begin{equation}
\hat{F}\left(\boldsymbol{x}^*,\boldsymbol{\alpha}^*\right)=\frac{F\left(\boldsymbol{x}^*,\boldsymbol{\alpha}^*\right)}{\min_{\boldsymbol{x}'\in\{0,1\}^N}F\left(\boldsymbol{x}',\boldsymbol{\alpha}_{\boldsymbol{x}'}\right)}
\end{equation}
where the numerator denotes the optimal offloading cost by DRTO algorithm, and the denominator denotes the optimal offloading cost by greedily enumerating all the $2^N$ offloading locations. We set the update interval $\Delta=64$. As we can see, within the first $5000$ time frames, the normalized offloading cost significantly fluctuates, indicating that the offloading policy has not fully converged. Finally, most of the normalized offloading cost are converged to $1$, only few frames slightly fluctuates above $1$ due to the rapid channel fading when inter-satellite handover occurs. In spite of this fluctuation, the DRTO algorithm can still achieve near-optimal offloading cost performance.
\subsection{The Performance of Offloading Cost}
Regarding to the offloading cost performance, we compare our DRTO algorithm with other five representative benchmarks to demonstrate its superiority:
\begin{itemize}
	\item Distributed Deep Learning-based Offloading (DDLO)\cite{Huang_DDLO_2018}. Multiple DNNs take the duplicated channel gain as input, then each DNN generates a candidate offloading location. Then, the optimal offloading location is selected with respect to the minimum offloading cost. In the comparison with DRTO, we assume that DDLO is composed of $N$ DNNs.
	\item Coordinate Descent (CD)\cite{Bi_CD_2018}. The CD algorithm is a traditional numerical optimization method, which iteratively swaps the offloading location of each STs that leads to the largest offloading cost decrement. The iteration stops when the offloading cost cannot be further decreased by swapping the offloading location.
	\item Enumeration. We enumerate all $2^N$ offloading location combinations and greedily select the best one.
	\item Pure TC Computing. The LEO access satellite forwards all the tasks to TC for executing.
	\item Pure SatEC Computing. The LEO access satellite locally execute all the tasks.
\end{itemize}
\begin{figure}[htbp]
	\centerline{\includegraphics[width = 0.45\textwidth]{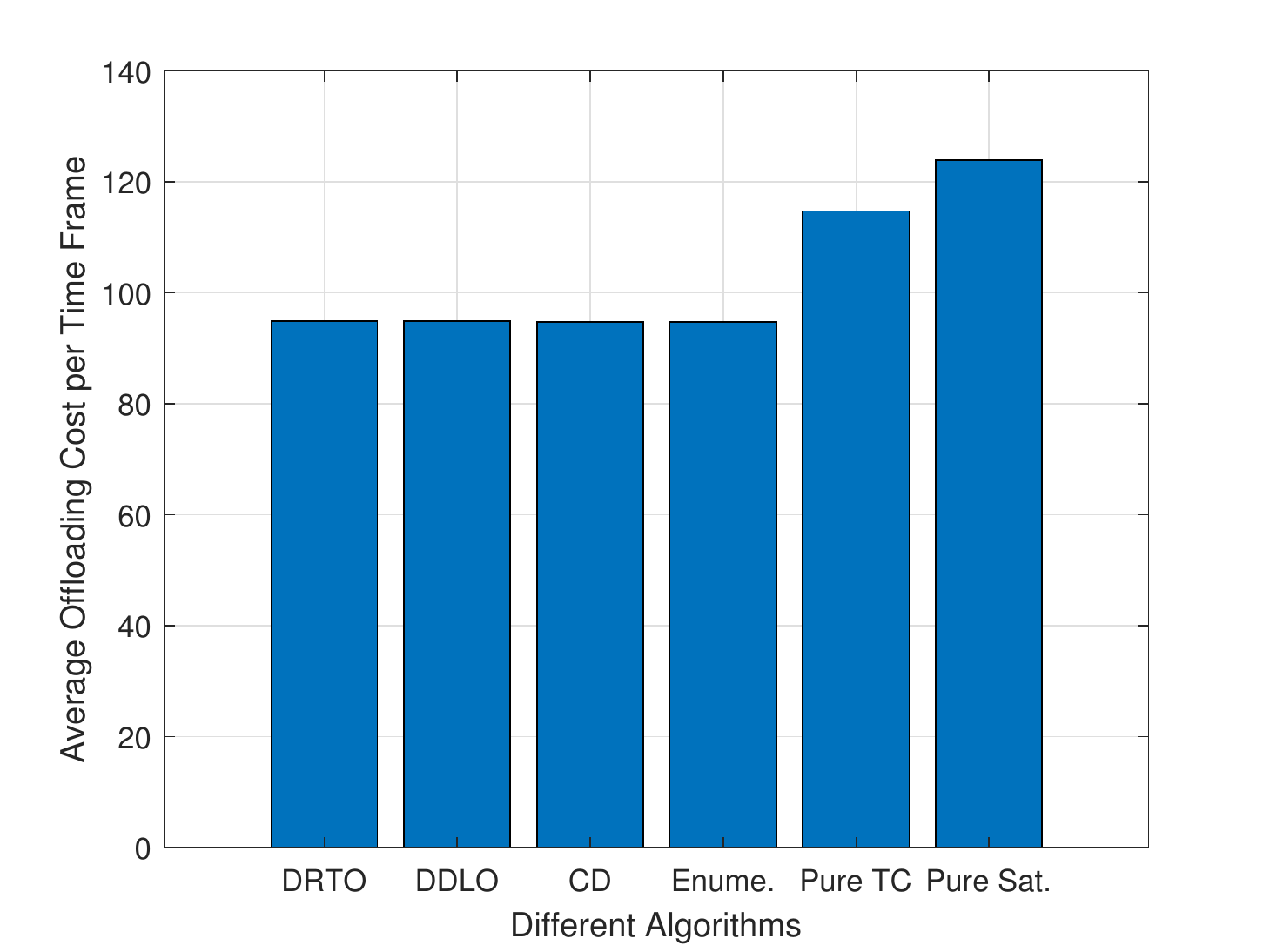}}
	\caption{Average Offloading Cost by Different Algorithms ($N=5$)}
	\label{offloading_cost_comparison}
\end{figure}

We consider $N=5$ STs attaching to the same access satellite. In Fig. \ref{offloading_cost_comparison}, we compare the performance of average offloading cost per time frame achieved by different offloading algorithms. As we can see, DRTO achieves similar performance as the greedy enumeration method, which verifies the optimality of DRTO. Since the optimal offloading location combination is unique, any other random combinations will lead to higher offloading cost. In addition, we see that DRTO achieves lower offloading cost with about 17.5\% and 23.6\% reduction compared to pure TC Computing and pure SatEC Computing methods, which indicates the necessity of cooperation between SatEC servers and TCs to provide satisfying computing service.
\subsection{The Performance of Runtime Consumption}
\begin{figure}[htbp]
	\centerline{\includegraphics[width =0.45\textwidth]{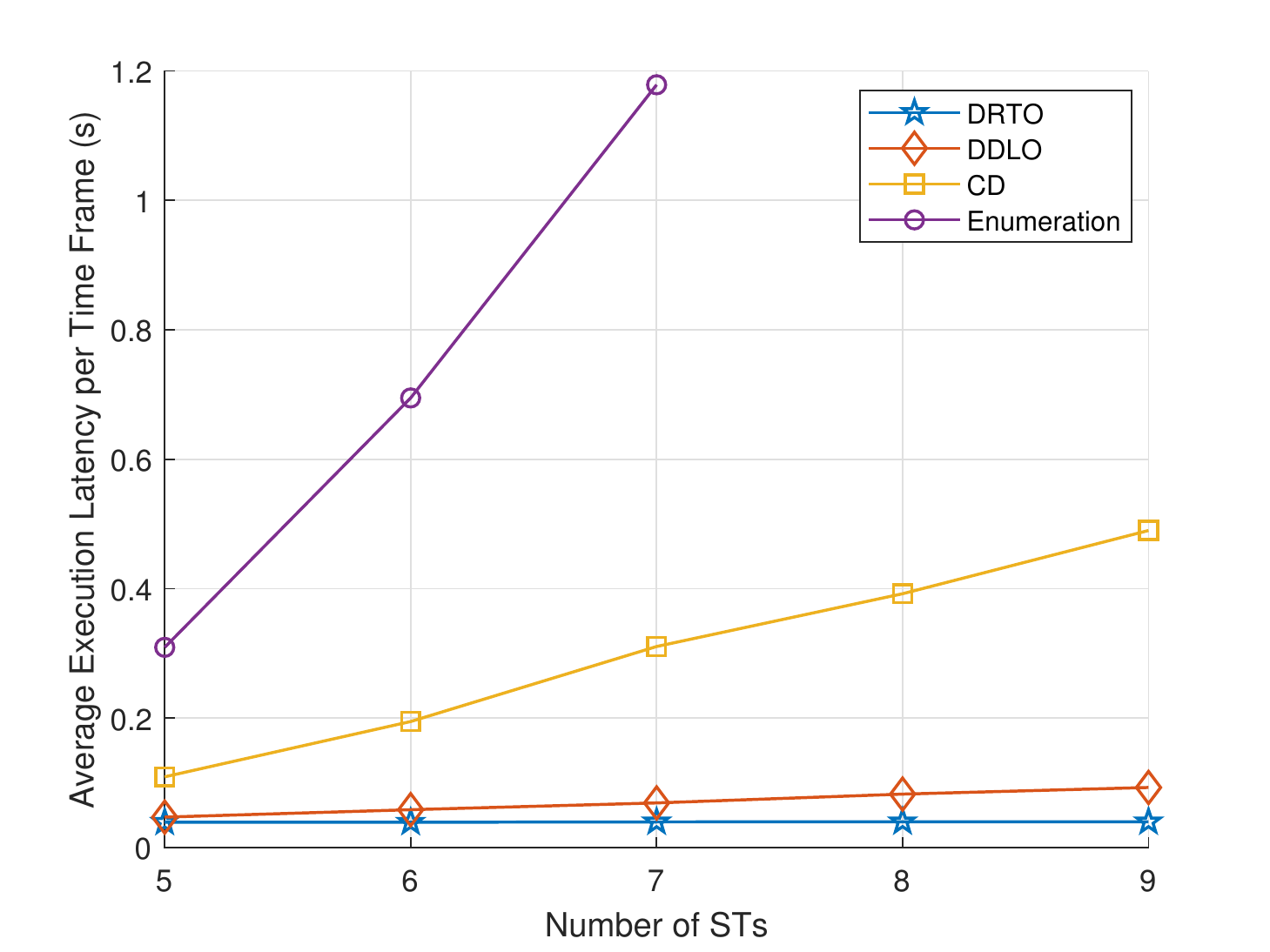}}
	\caption{Average Execution Latency by Different Algorithms}
	\label{exe_latency_comparison}
\end{figure}

Finally, we evaluate the runtime performance of DRTO. Since Pure TC Computing and Pure SatEC Computing are static, we compare DRTO with other three dynamic benchmarks. Specifically, we respectively record the total runtime consumption of different algorithms running on $30000$ time frames, and compute the average runtime per time frame. The runtime comparison is shown in Fig. \ref{exe_latency_comparison}. 

Although four dynamic algorithms achieve similiar offloading cost performance (in Fig. \ref{offloading_cost_comparison}), DRTO consumes the lowest runtime with about 42.6\%, 87.3\% and 96.6\% reduction comparing to DDLO, CD and Enumeration when $N = 7$. In addition, the runtime consumption of DRTO or DDLO does not explode when network scale increases. This is because that DNN can accurately fits the complex mapping from channel states to offloading location, compared with traditional CD or Enumeration methods, the action space of DRTO or DDLO is significantly reduced, resulting in much less iterations. In the comparison with DDLO, at the very beginning of learning process, the action space of DRTO is the same as that of DDLO. With the improvement of offloading policy, the number of quantized candidate offloading locations in DRTO is dynamically adjusted, thus the action space of DRTO is further reduced.

Actually, the channel coherent time is extremely short due to the high speed movement of satellites. DRTO can quickly generates offloading location and bandwidth allocation without compromising offloading cost performance, which better adapts to the fast channel fading in satellite-terrestrial edge computing networks.
\section{Conclusion}
\label{conclusion}
In this paper, we investigate the joint offloading location decision and bandwidth allocation problem in satellite-terrestrial edge computing networks, and propose DRTO algorithm to minimize the offloading cost based on current observed channel states. DRTO improves its offloading policy by learning from the past offloading experiences via reinforcement learning. To achieve faster convergence, we preserve order when generating candidate offloading locations and dynamically adjust $K$ during learning process. Simulation results show that our DRTO algorithm achieves near-optimal offloading cost performance as existing algorithms, but significantly reduces runtime consumption, making real-time offloading optimization truly viable under fast fading channel in satellite-terrestrial edge computing networks.

\section*{Acknowledgement}
This work was supported by the National Key Research and Development Program of China (No. 2017YFB0801900), Priority Research Program of Chinese Academy of Sciences (No. XDC02011000), Chinese National Key Laboratory of Science and Technology on Information System Security (No. 6142111190303), and Confidential Research Program (No. BMKY2018B17).

\bibliographystyle{IEEEtran}
\bibliography{bibliography}
\end{document}